\documentclass[final]{svjour3}
\usepackage{graphicx}
\usepackage{rotating}
\usepackage{amssymb}
\usepackage{mathptmx}
\usepackage[numbers]{natbib}
\makeatletter
\journalname{Journal of Low Temperature Physics}

\newcommand{\1}{\textsuperscript{1}} 
\newcommand{\2}{\textsuperscript{2}} 
\newcommand{\3}{\textsuperscript{3}} 
\newcommand{\4}{\textsuperscript{4}} 
\newcommand{\5}{\textsuperscript{5}} 

\begin{document}

\newcommand{\hdblarrow}{H\makebox[0.9ex][l]{$\downdownarrows$}-}
\title{HEMT-based 1K front-end electronics for the heat and ionization Ge CryoCube of the future R\textsc{icochet} CE$\nu$NS experiment}
\titlerunning{HEMT-based 1K front-end electronics}

\author{G. Baulieu\1, J. Billard\1, G. Bres\2, J-L Bret\2, D. Chaize\1, J. Colas\1, Q. Dong\5, O. Exshaw\2, C Guerin\1, S. Ferriol\1, J-B Filippini\1, M. De Jesus\1, Y. Jin\4, A. Juillard\1, J. Lamblin\3, H. Lattaud\1,  J. Minet\2, D. Misiak\1, A. Monfardini\2, F. Rarbi\3, T. Salagnac\1, L. Vagneron\1, and the R\textsc{icochet} Collaboration
}
\authorrunning{R\textsc{icochet} collaboration}


\institute{\1 Univ. Lyon, Univ. Lyon 1, CNRS/IN2P3, IP2I, F-69622, Villeurbanne, France\\
\email{alexandre.juillard@ipnl.in2p3.fr}\\
\2 Institut Néel, Université Grenoble-Alpes, CNRS/INP, 25 rue des Martyrs, Grenoble, France\\
\3 LPSC, Université Grenoble-Alpes, CNRS/IN2P3, 53, rue des Martyrs, Grenoble, France\\
\4 C2N, CNRS, Univ. Paris-Sud, Univ. Paris-Saclay, 91120 Palaiseau, France\\
\5 CRYOHEMT S.A.S 91400 Orsay, France
}

\maketitle

\begin{abstract}

The R\textsc{icochet} reactor neutrino observatory is planned to be installed at the Laue Langevin Institute (ILL) starting mid-2022. Its scientific goal is to perform a low-energy and high precision measurement of the coherent elastic neutrino-nucleus scattering (CE$\nu$NS) spectrum in order to explore exotic physics scenarios. R\textsc{icochet} will host two cryogenic detector arrays: the CryoCube (Ge target) and the Q-\textsc{array} (Zn target), operated at 10 mK.
The 1 kg Ge CryoCube will consist of 27 Ge crystals instrumented with NTD-Ge thermal sensors and charge collection electrodes for a simultaneous heat and ionization readout to reject the electromagnetic backgrounds (gamma, beta, x-rays).
We present the status of its front-end electronics. The first stage of amplification is made of High Electron Mobility Transistors (HEMT) developed by CNRS/C2N laboratory, optimized to achieve ultra-low noise performance at 1K with a dissipation as low as 15 $\mu$W per channel.
Our noise model predicts that 10 eV heat and 20 eV$_{ee}$ RMS baseline resolutions are feasible with a high dynamic range for the deposited energy (up to 10 MeV) thanks to loop amplification schemes. Such resolutions are mandatory to have a high discrimination power between nuclear and electron recoils at the lowest energies.


\end{abstract}

\section{Introduction}

High-purity cryogenic semiconductor detectors with simultaneous readout of phonon and ionization signals have been developed over the past three decades.
These developments were initially motivated by the direct search for Dark Matter in the form of Weakly Interacting Massive Particles (WIMPs) and more recently by the search of new physics with precise measurement of the coherent elastic neutrino-nucleus scattering (CE$\nu$NS) spectrum: in both cases the signal is a nuclear recoil while the background are electron recoils induced by $\alpha$, $\beta$ and $\gamma$ radioactivity and heat-only events.

The COHERENT collaboration reported the first observation of CE$\nu$NS from $\sim$ 30 MeV neutrinos  induced by the SNS spallation source \cite{Coherent}. At this high energy the scattering is not fully coherent and the influence of the new physics is expected to be more visible with neutrinos of few MeV. In germanium, the nuclear recoil energy spectrum for 5 MeV neutrinos is similar to that of a 2.7 GeV WIMP, this explains the important synergy between the development of low mass dark matter and CE$\nu$NS detectors.
The cryogenic Dark Matter detector community has therefore started to work on  projects such as R\textsc{icochet}, NUCLEUS and MINER \cite{Ricochet, Nucleus, Miner} to measure CE$\nu$NS from few MeV neutrinos produced by nuclear reactors.
 
R\textsc{icochet} will be installed at the Laue Langevin Institute \cite{ILL} in Grenoble (France). Two cryogenic detector arrays, the CryoCube (Ge target) and the Q-\textsc{array} (Zn target) will be running in a dilution fridge at 8.8 meter away from the 58 MW nuclear reactor, shielded by 20 cm of Pb and 35 cm of PE, and surrounded by an active plastic scintillator muon veto. The site provides an overburden of 15 m.w.e (meter water equivalent). More details on the R\textsc{icochet} setup are given in \cite{Ricochet}, the CryoCube is described in \cite{CryoCube}. With a $\sim$ 1 kg payload, a total of 13 CE$\nu$NS interactions are expected per day. The Ge detectors will be of two types: planar detectors with 2 charge collection electrodes (top and bottom) and Full Interditized Detectors adapted from the FID800 detectors developed by the EDELWEISS collaboration \cite{EDW_JINST} and fitted by 4 electrodes  (2 on the top, 2 on the bottom). All detectors will be equipped with 1 Ge-NTD heat sensor.
One of the main challenge of R\textsc{icochet} is to keep an active discrimination for nuclear recoils depositing energy in the detector as low as 50 eV. As the actual background at such low energy is still unknown and all cryogenic experiments exhibit huge excesses at low energy \cite{EXCESS} this approach is an important risk mitigation.
Due to the low ionization yield for nuclear recoils, the detection threshold is driven by the heat channel while the discrimination threshold is limited by the ionization channel resolution. A baseline resolution of $\sigma$ = 20 eV$_{ee}$ RMS (20eV-electron equivalent, ionization energy produced by an electronic recoil depositing 20 eV in the germanium detector and corresponding to about 7 electron-hole pairs) is needed to meet the goal. This is one order of magnitude below the best resolution reached by cryogenic Ge detectors with Si-JFET \cite{EDW_JINST}. The heat channel baseline resolution goal is 10 eV (RMS) which is less than a factor 2 of the best achieved resolution on such small Ge detector by the EDELWEISS collaboration \cite{EDW_Surf}.

In previous work \cite{HEMT_LTD18} we have shown that HEMTs developed by the CNRS/C2N laboratory  \cite{HEMT_YongAPL, HEMT_ICSICT}, and now commercially available at the CryoHEMT company \cite{CryoHEMT}, could be used as high impedance input devices and could meet the resolution goals with a total input (detector + cabling) capacitance of 20 pF. 
The noise measurements were made at 4 K with a dissipation of 100 $\mu$W. The extrapolation to the 150 channels needed for the CryoCube is not straightforward.
In this paper, we show that these HEMTs keep their performance at 1 K with power dissipation as low as 15 $\mu$W and exhibit a good reproducibility over about 100 HEMT tested. As a result the cold electronics could be placed at 1 K, very close to the detector stage to mitigate the cabling capacitance. In addition we show that precise gain and high linearity are obtained thanks to closed loop 300 K amplifiers with feedback on the input HEMT, first measurements at low temperature are presented.

\section{HEMTs performance and reproducibility at 4K and 1K }

\begin{figure} [!htbp]
\begin{center}
\includegraphics[%
  width=0.99\linewidth,
  keepaspectratio]{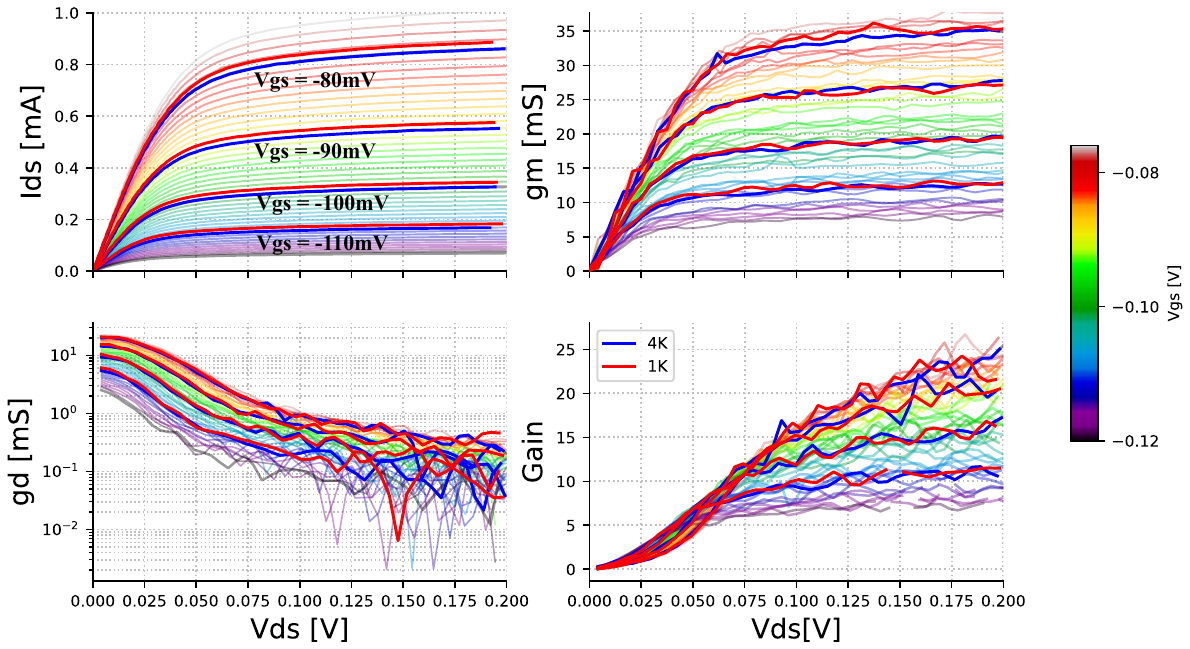}
\end{center}
\caption{{\it Top Left:} 
I$_{ds}$ vs V$_{ds}$ characteristics at 4K and 1K for a C$_{gs}$=100 pF HEMT device. Horizontal scales are the same for plots at top and bottom.
From top to bottom a $\delta$V$_{gs}$ = -1 mV step is applied between each curve for the 1K data, only 4 curves are shown for the 4K data (thick blue) to highlight the similarity with 1K curves. {\it Top Right and Bottom Left:} Extracted transconductances g$_{m}$ and output conductances g$_{d}$. 1K and 4K (thick blue) values are the same within the measurement precision. {\it Bottom Right:} Voltage gain A$_{V}$. 1K and 4K data (thick blue) are the same (Color figure online.)}
\label{IdeV}
\end{figure}

Commercially available HEMTs are aimed for GHz and 50 $\Omega$ matching circuits applications and suffer from high noise current and large low-frequency noise voltage  \cite{NoiseInHEMT}. The HEMTs developed by the CNRS/C2N laboratory are specially designed with optimizations of the hetero-structure growth with a relatively large spacer layer and the gate formation with a relatively large gate length in order to reduce their low-frequency current and voltage noise under cryogenic conditions.

HEMTs of various geometries have been developed , with input capacitance C$_{gs}$ varying from 2 pF to 230 pF. I$_{ds}$ vs V$_{ds}$ characteristics for various V$_{gs}$ have been first measured at both 4 K and 1 K up to a saturation current of $\sim$1 mA and from this data we have extracted the transconductance $g_{m} = \delta I_{ds}/\delta V_{gs}$  and the output conductance  $g_{d} = \delta I_{ds}/\delta V_{ds}$. The voltage gain A$_{V}$ is measured in a standard common source amplifier configuration with a 1 k$\Omega$ load resistance on the drain  \cite{HEMT_LTD18}.
Typical results are given in Fig.~\ref{IdeV} for a C$_{gs}$ = 100 pF device. The 1 K and 4 K characteristics are similar. The saturation decreases from 1 mA to 0.1 mA by varying V$_{gs}$  from -75 mV to -115 mV. The I$_{ds}$ vs V$_{ds}$ characteristics are shifted by only 1mV. The extracted g$_{m}$ and g$_{d}$ at 1 K and 4 K are compatible within the precision measurement as well as the gain A$_{V}$. This stability of HEMTs characteristics and performance at 1 K and 4 K are observed for all HEMTs geometries.

HEMT gain and power dissipation rise at large current, but there is  an upper limit given by the power handling capability of the refrigerator, a balance has to be
found as a reasonable gain must be achieved to mitigate the contraints on 300K electronics.

A good compromise is found with V$_{ds}$ = 100mV the saturation is well established for all V$_{gs}$. At  I$_{ds}$ =  0.15mA we have a low dissipation of 15 $\mu$W, a transconductance of 10 mS, an output conductance of 0.2 mS and a voltage gain of 8 (which corresponds to an intrinsic gain g$_{m}$/g$_{d}$ of $\sim$ 50). This working point will be used as a benchmark in the following.

The voltage and current noise have been characterized in details for all geometries using the methodology presented in  \cite{HEMT_YongAPL}. A summary on R\textsc{icochet} constraints is presented in \cite{HEMT_LTD18}. The voltage noise and current noise magnitudes vary as $1/\sqrt{C_{gs}}$ and $\sqrt{C_{gs}}$ respectively. As a result the 5 pF device is well adapted to ionization measurement with total input capacitance of about 20 pF, while the 100 pF and 230 pF devices are well fitted to Ge-NTD heat sensors up to few tens of M$\Omega$. With such input impedance the current noise contribution of the 5 pF is negligible with respect to its voltage noise and the current noise contribution of the 100 pF and 230 pF devices are negligible with respect to the thermal noise of the Ge-NTD sensor.

\begin{figure} [!htbp]
\begin{center}
\includegraphics[%
  width=0.495\linewidth,
  keepaspectratio]{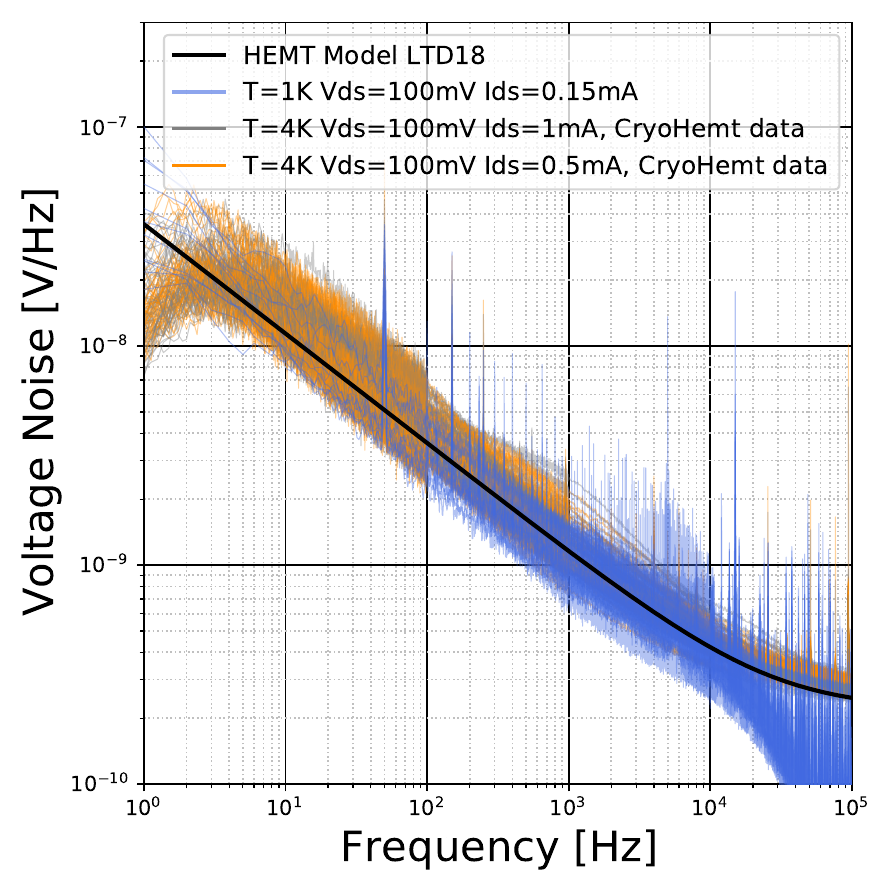}
  \includegraphics[%
  width=0.495\linewidth,
  keepaspectratio]{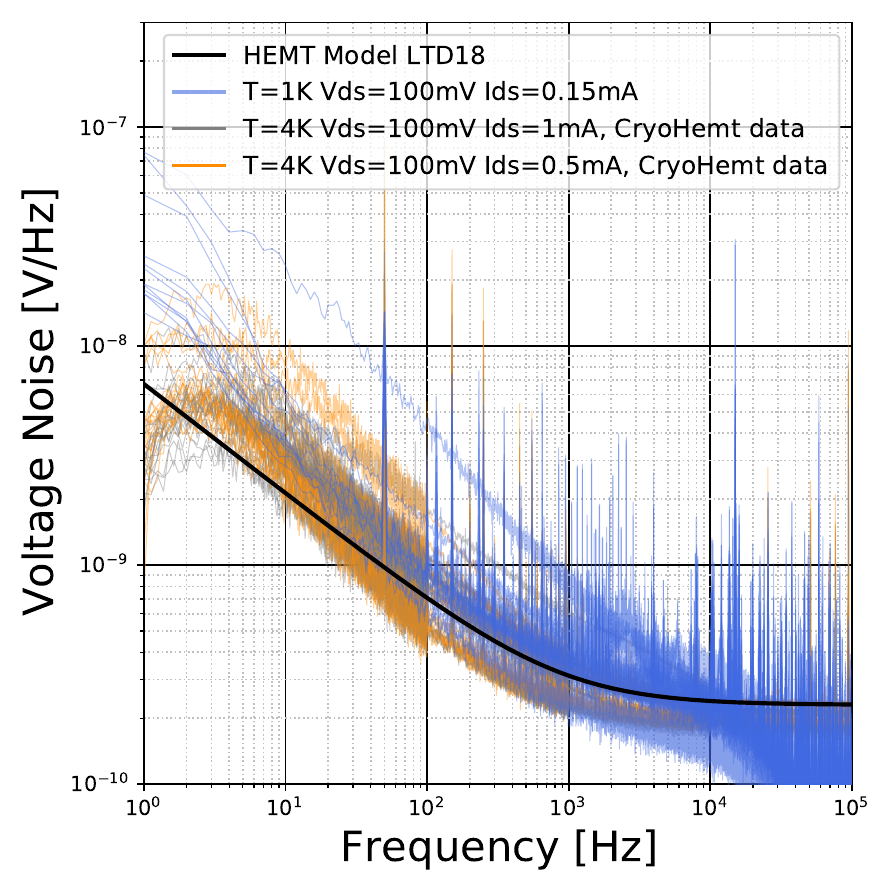}
\end{center}
\caption {Voltage noise Power Spectral Density at 3 working points for T, V$_{ds}$ and  I$_{ds}$ at 4K, 100 mV, 1 mA; 4K, 100 mV, 0.5 mA and 1K, 100 mV, 0.15 mA respectively. The noise PSD at 4K are provided by CryoHEMT and are not corrected from a gain variation below 10 Hz due to the bias output impedance. The noise PSD at 1K have been measured at Lyon with a gain stable down to 2 Hz.
{\it Left:} 
C$_{gs}$ = 5 pF HEMT device.
{\it Right:} 
C$_{gs}$ = 100 pF and C$_{gs}$ = 230 pF HEMT devices. 
Thick black curve is the targeted noise expected from \cite{HEMT_LTD18}. No anti-aliasing filter is applied on this model.
(Color figure online.)}
\label{en}
\end{figure}

Voltage noise measurements of 65 C$_{gs}$ = 5 pF devices and 22 C$_{gs}$ = 100-230 pF devices have been provided by the CNRS/C2N laboratory. The noise is measured at 4.2 K in an He dewar with  V$_{ds}$ = 100 mV and I$_{ds}$ = 0.5 and 1 mA. 13 of the 100-230 pF HEMTs and 12 of the 5 pF have then been measured at 1K at the low dissipation working point benchmark in the IP2I dilution fridge for comparison. Results are shown on Fig.~\ref{en}. The noise has been measured in common source configuration.

The results for the 5 pF HEMTs are shown on Fig.~\ref{en} ({\it Left}). The noise level is the same for the 3 configurations (4K-1mA, 4K-0.5mA, 1K-0.15mA) without any outliers. The voltage noise is compared to the 5 pF voltage noise model developed for R\textsc{icochet} \cite{HEMT_LTD18}. For a given input impedance, the detector input sensitivity is known and the optimal filtering formalism can be used to extract the expected baseline resolution \cite{HEMT_LTD18}. The model gives a RMS baseline resolution of $\sigma=$20 eV$_{ee}$ for a total detector + cabling capacitance of 20 pF. All baseline resolutions deduced from the 1 K data are in the 16 to 25 eV$_{ee}$ range and 80 $\%$ lie between 18 to 22 eV$_{ee}$. With a proper preselection of HEMTs and additional measurement at 1K, the R\textsc{icochet} objectives to have up to 120 5 pF HEMTs with voltage noise as expected by the model  \cite{HEMT_LTD18} seems reachable by mid-2022.

The noise measurement at the same working point for the 100 pF and 230 pF HEMTs are shown on Fig.~\ref{en} ({\it Right}). The spread between the measurements is larger compared to the 5 pF HEMTs, which could be due to the larger gate surface or a worse preselection of the HEMTs by CryoHEMT. The extreme outlier in the 1 K dataset is not explained as it passed the same quality cut as the other measurements. Unlike the ionization signal, the detector heat sensor impedance is not sufficient to predict the heat sensitivity. To extract baseline resolutions from the noise PSD we thus consider a typical pulse shape for the heat signal and normalise the result to an arbitrary sensitivity. 
The $\sigma=$ 18 eV RMS resolution obtained in \cite{EDW_Surf} with Si-JFET based electronics have been obtained with a Ge detector of same mass as planed for R\textsc{icochet} and a sensitivity of $\sim $ 1.5 $\mu$V/keV. 
With this sensitivity the heat baseline resolution predicted with the 1 K data are then in the 16-20 eV range.
In fact most of the HEMTs tested have a noise level higher than expected if compared to the model developed in \cite{HEMT_LTD18} at the origin of the 10 eV goal. To meet the 10 eV goal R\textsc{icochet} is developing detector with higher sensitivity and works are in progress in collaboration with C2N and CryoHEMT to have a batch of thirty 230 pF HEMT all with the expected noise levels.

\section{HEMT-based charge amplifier for the ionization channel}

 Proper design of the cold and 300 K second stage of the amplification is required to optimize noise performances.
Previous work \cite{BerkeleyNIM} has demonstrated a 91 eV$_{ee}$ RMS baseline ionization resolution with a HEMT-based cryogenic charge amplifier coupled to a 240 g, 150 pF CDMS-II cryogenic germanium detector, nearly matching the expected predictions. However, the complexity required to implement a HEMT-based fully cryogenic charge amplifier for the hundred of ionization channels present in R\textsc{icochet} requires to explore amplifier topologies where the use of cryogenic HEMTs is limited to only the input transistor. 
The ionization channel also needs to show a wide dynamic range between the threshold energy up to tens of MeV expected from high energy gammas and muons. The gain must be stable at the $\%$ level to allow for a good reconstruction of the charge signal.
For this purpose, the basic solution is a feedback topology that exploits the high open loop gain of a low noise op amp. The feedback can be on the HEMT gate or on its source. Main issues can arise from possible oscillations and instabilities induced by the long cabling between the 1 K cold stage and the 300K electronics, which adds a phase shift that could compromise stability.

For R\textsc{icochet} the 300 K stage will be more than 2 m away from the 1 K stage. However the constraints on the bandwidth are low, it has to match the detector - muon veto coincidence window, a limited 40 kHz bandwidth is sufficient according to our simulation \cite{HDRJulien}. There are no contraints on the low frequency cut as long as the resolution goal is met; since the signal/noise ratio is high at low frequency, a cutoff at few Hz is expected.

\begin{figure} 
\begin{center}
\includegraphics[%
  width=0.999\linewidth,
  keepaspectratio]{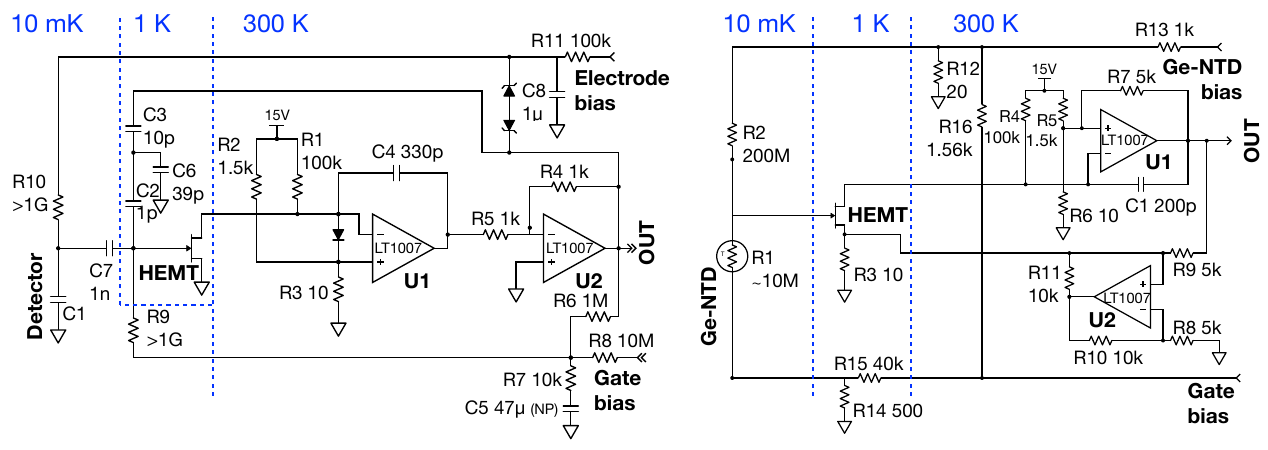}
\end{center}
\caption{
Schematics of the charge ({\it Left}) and heat ({\it Right}) amplifier. Values of resistances and capacitances are optimized for a 20 pF detector + cabling input capacitance and a GeNTD heat sensor biased at about 10 M$\Omega$. 
See text for details. (Color figure online.)}
\label{AmpliCharge}
\end{figure}

Fig.\ref{AmpliCharge} ({\it Left}) shows the schematic of the first version of the R\textsc{icochet} charge amplifier. 
DC biases for both detector and electronics are made using low noise (6 nV/$\sqrt{Hz}$) DACs (not shown on the Figure).
The source is grounded, V$_d$ and I$_{ds}$ are fixed by the op amp U1 and the R1, R2, R3 resistor network to 100 mV and 0.15 mA respectively. 
The gate is biased thanks to a DC feedback loop of the global amp output through the high value R9 resistor ($>1 G\Omega$ at 10 mK to minimize noise). The output voltage offset can be adjusted through R8. The bias noises are attenuated by the filtering provided by R6-R7-C5 and R8-R7-C5 respectively.
The R6/R7=100 divider in the loop provides a boost to R9 which has then an effective value multiplied by 100 up to the 0.34 Hz low frequency cutoff given by R7 and C5.

The AC response of the amplifier can be described as follows.
High open loop gain is given by the HEMT cascaded with the low noise op amps U1 and U2 (LT1007 from Linear Technology).
The transient current at the drain and negative input of U1 is g$_{m}$ * $\delta$V$_{gs}$. As long as the U1 open loop gain is high this current is forced to pass through C4. Op amp U2 is an inverter to assure negative feedback..
The open loop gain of the HEMT + U1 + U2 chain is thus:
\begin{equation}
\label{olgain}
G_{OL} =  \frac{- g_{m}} {2\pi f \cdot C4}\approx \frac{- 10\cdot 10^6}{f}
\end{equation}

The U2 output is then connected to the HEMT input through the AC feedback provided by the C2-C3-C6 capacitor network, which acts as an effective feedback capacitance $C_{fb} = C2/(1+C2/C3+C6/C3) = C2/5$ connecting the U2 output and the HEMT input. The Closed Loop gain is :

\begin{equation}
\label{clgain}
G_{cl} = \frac{- \alpha}{1+\alpha\beta} = \frac{- \alpha}{1+ \alpha \left(\frac{C_{fb}}{C1 + C_{fb}}\right)} \approx \frac{- C1}{C_{fb}}\approx \frac{5\cdot C1}{C2}\approx - 100
\end{equation}
where $\alpha$ is the absolute value of the open loop gain, $\beta$ is the feedback factor, and $C1$ is the detector + cabling capacitance. The approximation holds as long as $\alpha\beta \gg 1$ and $C1 \gg C_{fb}$.

\begin{figure} 
\begin{center}
\includegraphics[%
 width=0.325\linewidth,
 keepaspectratio]{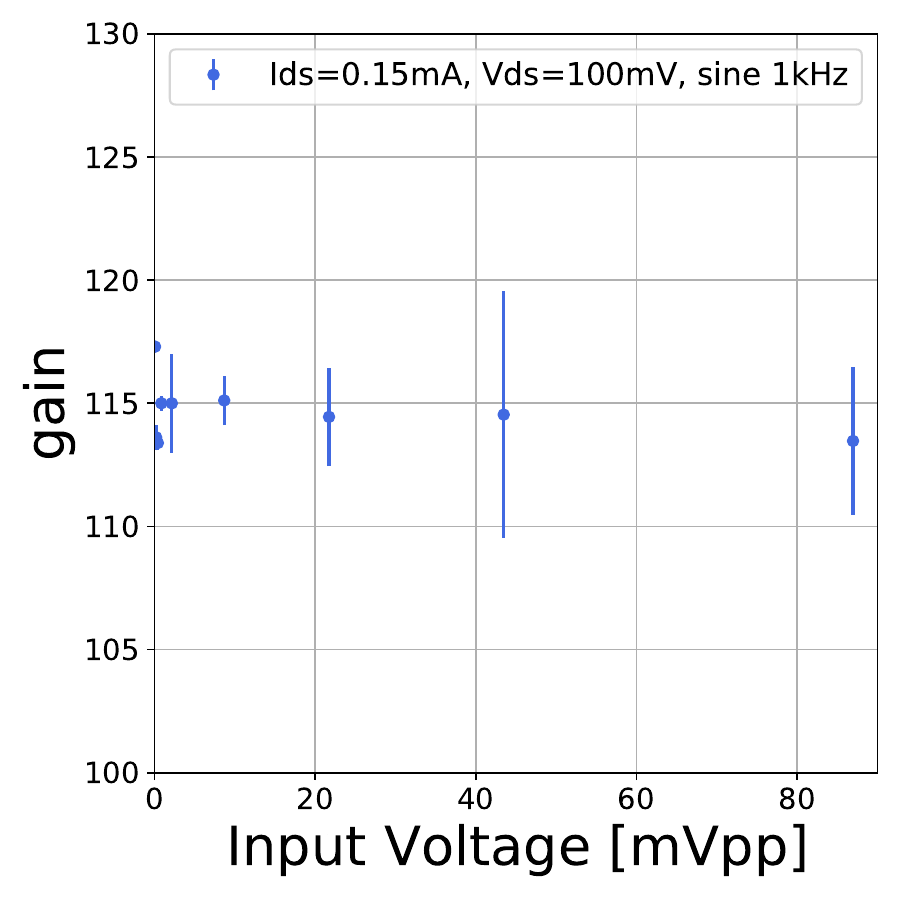}
 \includegraphics[%
 width=0.325\linewidth,
 keepaspectratio]{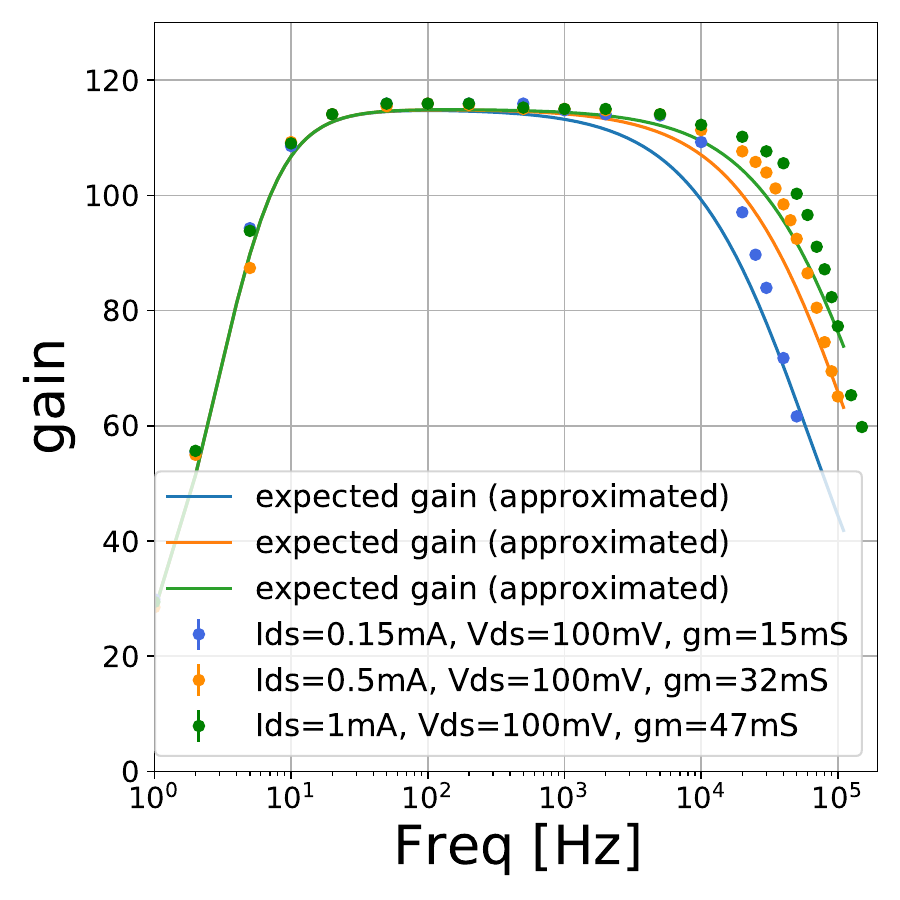}
 \includegraphics[%
 width=0.325\linewidth,
 keepaspectratio]{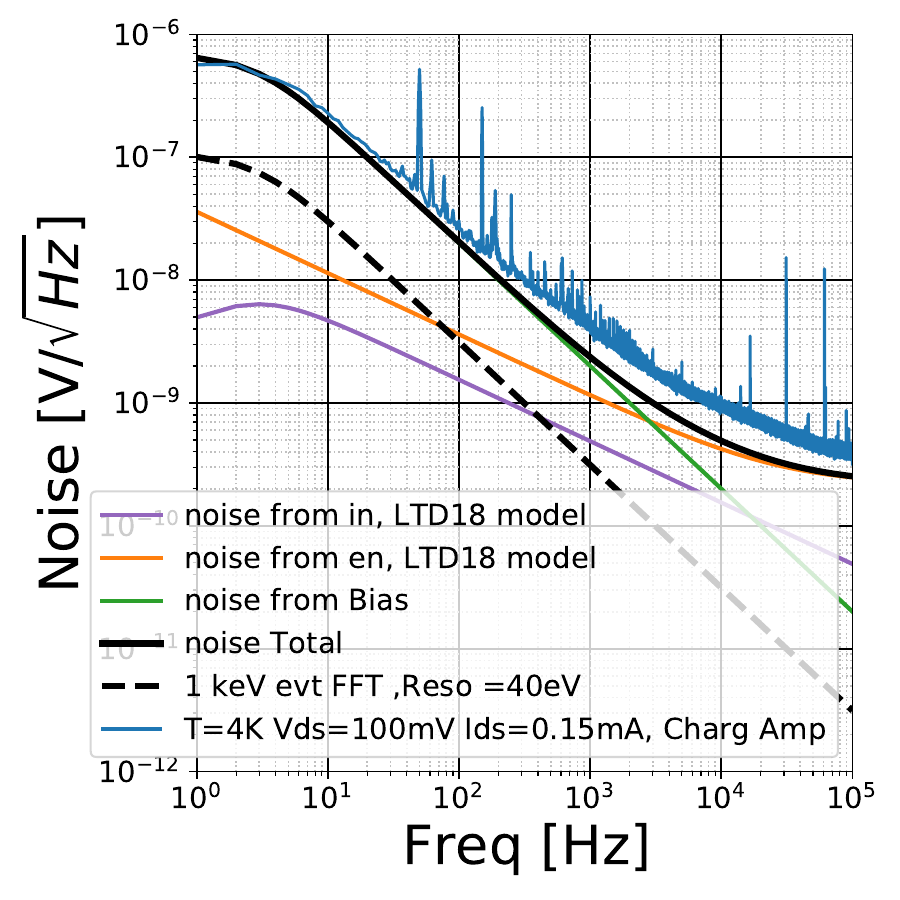}
 \includegraphics[%
 width=0.325\linewidth,
 keepaspectratio]{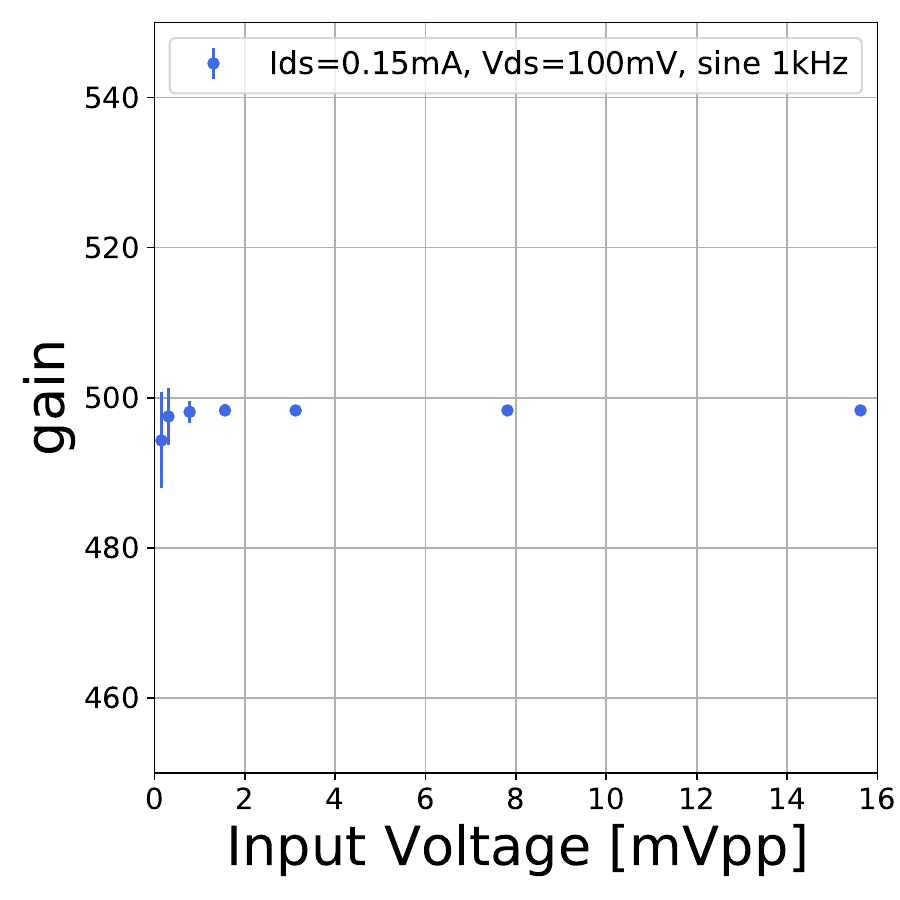}
 \includegraphics[%
 width=0.325\linewidth,
 keepaspectratio]{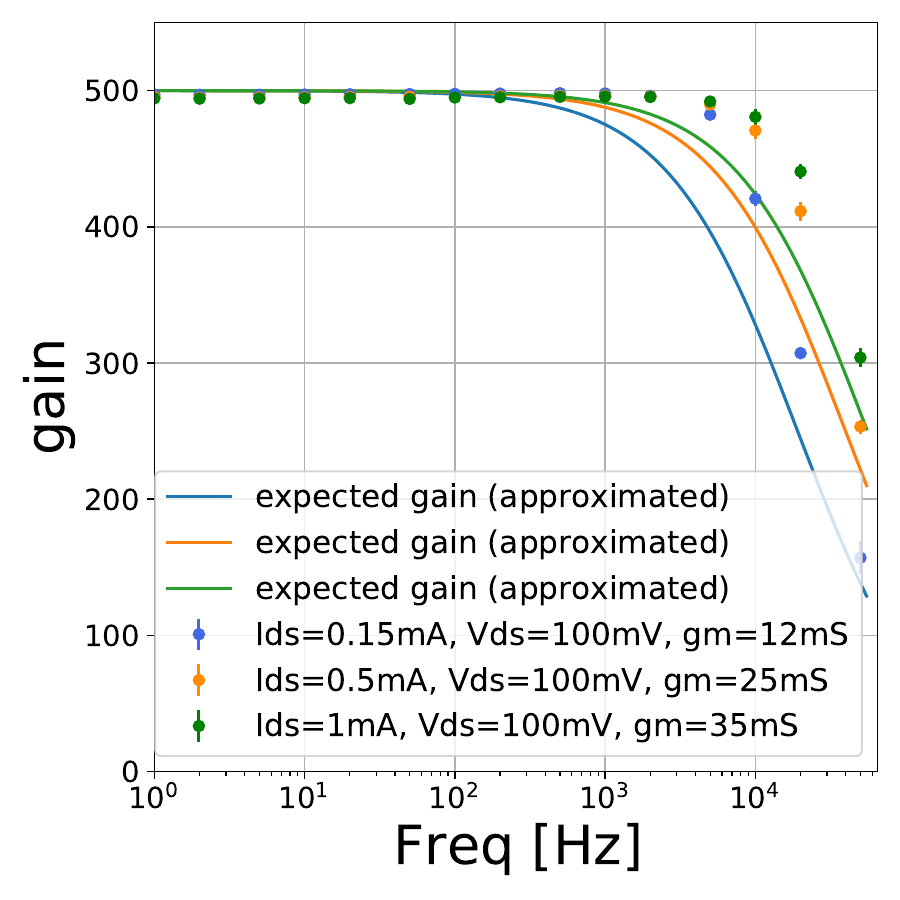}
 \includegraphics[%
 width=0.325\linewidth,
 keepaspectratio]{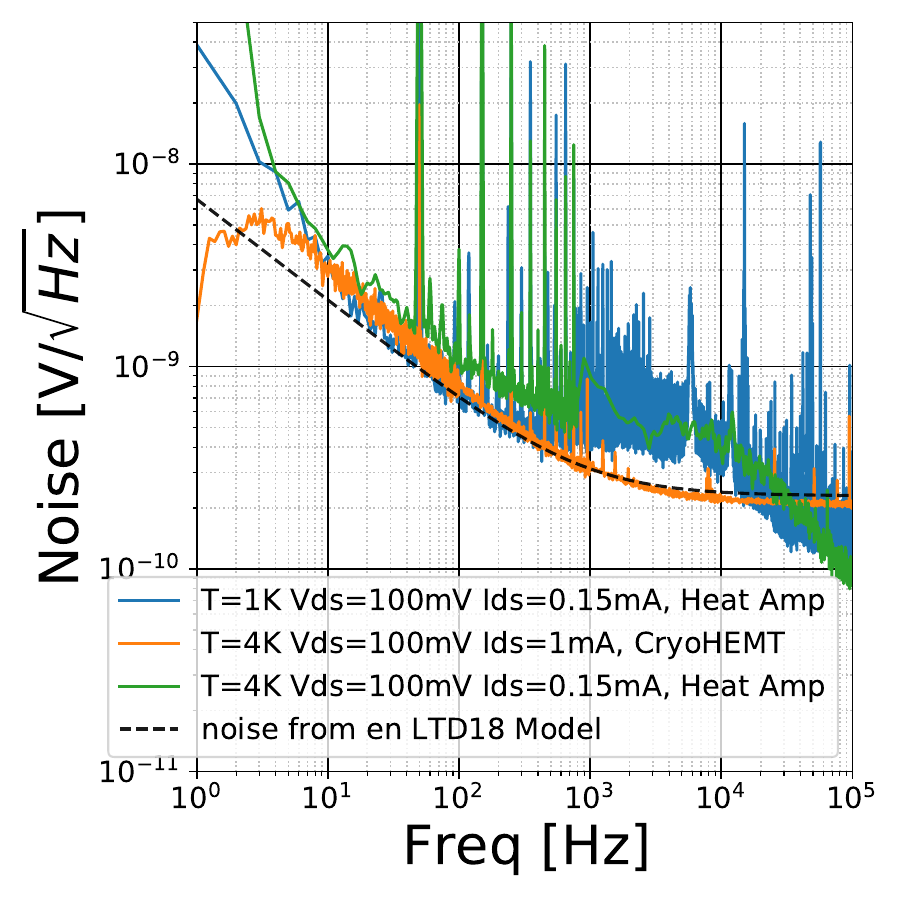}
\end{center}
\caption {
{\it Top Panels:} Performance of the charge amplifier at 4.2K with a 2 G$\Omega$ feedback resistance. \qquad \qquad
{\it Bottom Panels:} Performance of the voltage heat amplifier at 4.2K and 1K. NTD dummy resistance shorted (R1 = 0$\Omega$).
{\it Left:} 
1 kHz sine injected at the amp input to measure the linearity of the gain up to 85 mV (eq. to about 32 MeV) for the charge amplifier and 16 mV (eq. to about 15 MeV) for the heat amplifier.
{\it Center:} 
Bandwidth measurement at 3 working points (V$_{ds}$ = 100 mV and I$_{ds}$ = 0.15, 0.5 and 1 mA).
{\it Top Right:} 
Noise PSD of the charge amplifier at 4K 100 mV 0.15 mA. Various contributions to the noise, as expected by \cite{HEMT_LTD18}.
{\it Bottom Right:} 
Noise PSD of the heat amplifier at 4K, 100 mV, 0.15 mA and 1K, 100 mV, 0.15 mA are compared to the voltage noise of the same HEMT provided by CryoHEMT and the 100 pF HEMT noise model from  \cite{HEMT_LTD18}
(Color figure online.)}
\label{ResultChargeV1}
\end{figure}

The first prototype of this charge amplifier has been tested à 4.2 K in an helium dewar. R9 was a 1 G$\Omega$ resistor (HVC0805 from Ohmite) with a measured value of  $\sim$ 2 G$\Omega$ at 4.2K.
Results are presented on Fig.~\ref{ResultChargeV1} ({\it Top Panels}). 

The linearity was tested by injecting a known sinusoidal (1 kHz) voltage at the input via a 1 pF capacitor. The amplitude was varied up to 85 mV, which corresponds to about 32 MeV$_{ee}$ on a 20 pF input capacitance. Fig.~\ref{ResultChargeV1} ({\it Top Left}) shows that the linearity is excellent, with a stable gain in all the dynamic up to the output saturation occurring at 15 V.
The feedback loop is working nicely to stabilize the gain, getting rid especially of the important transconductance variation at large amplitude. The gain is measured to be 115 for an expected value of 100. This may come from stray capacitance in the test PCB and variation of the 20 pF dummy detector capacitance from 300 K to 4K even if COG-NPO material was adopted as the dielectric.

The bandwidth was measured by varying the frequency of a 1mV$_{pp}$ sine wave.
Three working points have been explored with V$_{ds}$ = 100 mV and I$_{ds}$ = 1 mA, 0.5 mA and 0.15 mA to measure the influence of the transconductance. Data are compared to the approximated gain expected from Eq.\ref{olgain} and Eq.\ref{clgain} on Fig.~\ref{ResultChargeV1} ({\it Top Center}). The low frequency is well explained by the DC feedback frequency cut given by R9*100 and C2/5 (4 Hz for 2 G$\Omega$ and 0.2 pF). At frequencies above 10 kHz we measured a gain slightly larger than expected for the 3 working points. This is due to the approximation done in Eq.\ref{olgain} where the gain variation of U1 with frequency is not included. The measured high frequency cut varies from 35 kHz to 80 kHz, well within the R\textsc{icochet} goals.

A stream of few minutes of output data have been recorded at a sampling rate of 200 kHz. 
The data are first corrected from the gain relative to the amp input and the noise PSDs are extracted after few basic quality cuts. Fig.~\ref{ResultChargeV1} ({\it Top Right}) shows that the noise is dominated by the thermal noise of R9 at 4.2 K. With this high thermal noise the expected baseline resolution is 40 eV$_{ee}$ RMS. 

Future work will consist of tests at 1 K with a dummy detector and an actual 20 pF detector at 20 mK few cm away from the HEMT input to limit the stray capacitance. The feedback resistor R9 will be replaced by an active reset scheme based on low capacitance HEMT, CMOS ASICs or AsGa diode to reach the 20 eV$_{ee}$ needed for the R\textsc{icochet} CryoCube.

\section{HEMT based voltage amplifier for the heat channel}

The heat amplifier shown on Fig.~\ref{AmpliCharge} ({\it Right}) uses some of the same features of the charge amplifier. The second stage of amplification at 300 K is based on the same low noise op amp to provide the needed high open loop gain thanks to its feedback capacitance C1.
The HEMT is biased in the same way. V$_d$ and I$_{ds}$ are kept constant and fixed by R4, R5, R6 and op amp U1.
The NTD is DC coupled to the HEMT input. This allows a direct bias of the gate through the R14 and R15 resistors and the NTD itself. The R12-R13-R16 network is added to keep the NTD bias constant while the gate bias varies.
The HEMT source is self-biased with the current flowing through the low value resistor R3.
The HEMT is used as a differential device, the amplification loop is closed by connecting the output of U1 to the HEMT source. As the feedback is on the source there is no need of inverter compared to the charge amplifier.
Thanks to this feedback V$_s$ follows V$_g$ to keep I$_{ds}$ constant.
The amplification scheme is equivalent to a non inverting amplifier with the HEMT gate at the positive input and the source at the negative input. In the bandwidth of the amplifier where the open loop gain is high enough the closed loop voltage gain  is 1+R9/R3 = 501.
Two solutions are added to improve linearity and gain stability: R7-R6 adds the small drain variation on the feedback loop with the same gain, so that V$_{ds}$ is always maintained  constant. The cryogenic cabling can have a large resistance and modify slightly the gain due to the addition of resistance in series with R9. To avoid this effect U2 is mounted as a Howland source: the current flowing through R3 is independent on the resistance between R9 and R3.

The first prototype of this heat amplifier has been tested both at 4.2 K in an Helium dewar and at 1K in a dilution fridge. The R1 dummy NTD resistance was shorted in order to measure only the voltage noise of the HEMT.
As for the charge amplifier the data taking focus on gain linearity, bandwidth and noise measurements.
The results are presented on Fig.~\ref{ResultChargeV1} ({\it Lower Pannels}). 

The linearity was tested by injecting a 1 kHz sine voltage at the HEMT input via the Ge-NTD bias line. The amplitude was varied up to 16 mV, which corresponds to about 15 MeV for a typical detector sensitivity. Fig.~\ref{ResultChargeV1} ({\it bottom Left}) shows that the linearity is excellent, with a gain stable over the entire dynamic range up to the output saturation.
As for the charge amplifier the feedback loop is working nicely to stabilize the gain. The gain is measured to be 498 for an expected value of 501. 

The bandwidth was measured in a same way as for the charge sensitive amplifier at the same working points. Data are compared to the approximated gain expected from the open loop and closed loop gain on Fig.~\ref{ResultChargeV1} ({\it Bottom Center}). 
The amplifier is DC coupled to the Ge-NTD, thus there is no low frequency roll-off. For the same reason as for the charge sensitive amplifier, at frequency above few kHz we measured a gain higher than expected for the 3 working points. The measured high frequency cut vary from 15 kHz to 40 kHz, well within the R\textsc{icochet} goals.

The voltage noise of the amplifier has been measured with the dummy Ge-NTD sensor shorted (R1 = 0$\Omega$). Fig.~\ref{ResultChargeV1} ({\it Bottom Right}) shows that the noise plateau of the 1K measurement in the few 100 Hz- few kHz frequency range is $\sim$ 0.5nV/$\sqrt{Hz}$. This is bit higher than the HEMT intrinsic voltage noise but the thermal noise of the Ge-NTD (few M$\Omega$ at 10 mK) will be higher anyway. At lower frequencies the 1 K data follow the intrinsic HEMT voltage noise up to few Hz, demonstrating that the voltage noise of the heat amplifier is adding a negligible noise contribution. 

Tests are ongoing with the input resistor at high impedance to measure the current noise contribution of the amplifier and with a detector prototype to demonstrate the 10 eV (RMS) heat baseline resolution.

 \section{Conclusion : }

The development of HEMT-based amplifiers for charge and heat channels represents a clear path to achieve the energy resolution, linearity and bandwidth requirements of R\textsc{icochet} for pursuing new physics in the CE$\nu$NS sector. Future work will address the issues of very low capacitance cabling, active reset schemes on the ionization channels and heat sensitivity improvement needed to reach the 20 eV$_{ee}$ ionization and 10 eV heat baseline (RMS) resolution goals.

\begin{acknowledgements}
This work was supported by the LabEx Lyon Institute of Origins (ANR-10-LABX-0066) of the Universite de Lyon in the framework "Investissements d'Avenir" (ANR-11-IDEX-00007) and has received funding from the European Research Council (ERC) under the European Union's Horizon 2020 research and innovation program under Grant Agreement ERC-StG-CENNS 803079.
We thank the EDELWEISS collaboration for the help and useful discussion in performing this work.
\end{acknowledgements}

\pagebreak


\begin{thebibliography}{99}

\bibitem{Coherent}
Observation of Coherent Elastic Neutrino-Nucleus Scattering, D. Akimov et al. (COHERENT collaboration), Science 357 (2017) 1123-1126, https://doi.org/10.1126/science.aao0990

\bibitem{Ricochet}
S. Hertel et al, Ricochet Progress and Status, J. Low. Temp. Phys. This Special Issue (2021)

\bibitem{Nucleus}
Exploring CEvNS with NUCLEUS at the Chooz Nuclear Power Plant, G. Angloher et al. (NUCLEUS collaboration), https://arxiv.org/abs/1905.10258

\bibitem{Miner}
Background studies for the MINER Coherent Neutrino Scattering reactor experiment
Author links open overlay panel, G.Agnolet et al., Nucl. Instrum. Meth. A853 (2017) 53. https://doi.org/10.1016/j.nima.2017.02.024

\bibitem{ILL}
https://www.ill.eu for general information on ILL.\\
https://www.ill.eu/reactor-and-safety/high-flux-reactor/technical-characteristics for reactor information.

\bibitem{CryoCube}
T. Salagnac et al, Optimization and performance of the CryoCube detector for the future R\textsc{icochet} low-energy neutrino experiment, J. Low. Temp. Phys. This Special Issue (2021)

\bibitem{EXCESS}
EXCESS Workshop June 2021
https://indico.cern.ch/event/1013203/
summary paper to be published

\bibitem{EDW_JINST}
E. Armengaud, et al., Performance of the EDELWEISS-III experiment for direct dark matter searches, J. Instrum. 12 (08) (2017) P08010. https://doi.org/10.1088/1748-0221/12/08/P08010

\bibitem{EDW_Surf}
Searching for low-mass dark matter particles with a massive Ge bolometer operated above ground, E. Armengaud et al. (EDELWEISS Collaboration), Phys. Rev D 99, 082003. https://doi.org/10.1103/PhysRevD.99.082003

 \bibitem{HEMT_LTD18}
A. Juillard et al, Low-Noise HEMTs for Coherent Elastic Neutrino Scattering and Low-Mass Dark Matter Cryogenic Semiconductor Detectors, J. Low. Temp. Phys. 199(3):793-806, May 2020.
https://doi.org/10.1007/s10909-019-02269-5

\bibitem{HEMT_YongAPL}
Q. Dong, et al., Ultra-low noise high electron mobility transistors for high-impedance and low-frequency deep cryogenic readout electronics, Appl. Phys. Lett. 105 (2014) 013504. http://dx.doi.org/10.1063/1.4887368.

\bibitem{HEMT_ICSICT}
Y. Jin et al.,  Ultra-low noise HEMTs for deep cryogenic low-frequency and high-impedance readout electronics, 12th IEEE International Conference on Solid-State and Integrated Circuit Technology (ICSICT) 2014. http://dx.doi.org/10.1109/ICSICT.2014.7021379.

 \bibitem{CryoHEMT}
 http://cryohemt.com/

\bibitem{BerkeleyNIM}
A. Phipps et al., A HEMT-based cryogenic charge amplifier with sub-100 eVee ionization resolution for massive semiconductor dark matter detectors, Nuclear Inst. and Methods in Physics Research, A 940 (2019) 181. https://doi.org/10.1016/j.nima.2019.06.022

 \bibitem{HDRJulien}
https://tel.archives-ouvertes.fr/tel-03259707
Searching for Dark Matter and New Physics in the Neutrino sector with Cryogenic detectors, Julien Billard, Habilitation à Diriger des Recherches. Jan 2021

 \bibitem{NoiseInHEMT}
R. Plana, L. Escotte, O. Llopis, H. Amine, T. Parra, M. Gayral, and J. Graffeuil, Noise in AlGaAs/InGaAs/GaAs pseudomorphic HEMTs from 10 Hz to 18 GHz, IEEE Trans. Electron Devices 40, 852 (1993). https://doi.org/10.1109/16.210190



 

 

 

 



\end{thebibliography}
\end{document}